\documentclass[aps,prl,reprint,superscriptaddress,nobibnotes]{revtex4-2}

\usepackage{amsmath}
\usepackage{amssymb}
\usepackage{mathrsfs}
\usepackage{array}
\usepackage{float}
\usepackage{soul}
\usepackage{graphicx}
\usepackage{xcolor}
\definecolor{brightcyan}{RGB}{0, 255, 255}
\sethlcolor{brightcyan}

\begin{document}





\title{Neural Learning as a Game Induced by Spike-Timing-Dependent Plasticity}


\author{Xinhao Fan}
\email{Contact author: xfan20@jhu.edu}
\affiliation{The Solomon H. Snyder Department of Neuroscience, Johns Hopkins University, Baltimore, MD 21218, USA.}

\author{Shreesh P. Mysore}
\email{Contact author: mysore@jhu.edu}
\affiliation{Department of Psychological and Brain Sciences, Johns Hopkins University, Baltimore, MD 21218, USA.}
\affiliation{The Solomon H. Snyder Department of Neuroscience, Johns Hopkins University, Baltimore, MD 21218, USA.}
\affiliation{Kavli Neuroscience Discovery Institute, Johns Hopkins University, Baltimore, MD 21218, USA.}





\date{\today}

\begin{abstract}

A general framework for inferring the computational role of spike-timing-dependent plasticity (STDP) does not currently exist. Here, we develop a game-theoretic description for a canonical, convergent neural circuit with general STDP and postsynaptic-potential (PSP) kernels. Parity-matched STDP–PSP interactions induce a potential game among presynaptic neurons, whereas parity-mismatched interactions induce a zero-sum game. These components, respectively, implement contrastive PCA and flow selection; their weighted interplay determines learning dynamics and computation for arbitrary STDP rules.  


\end{abstract}

\maketitle



A central problem in biological learning is understanding how local plasticity rules collectively give rise to global computation. Among the various plasticity mechanisms discovered over the past several decades, spike-timing-dependent plasticity (STDP) is one of the most extensively studied and experimentally validated \cite{dan2004spike, caporale2008spike}. It has been implicated in a wide range of learning-related functions in neural systems \cite{feldman2000timing, tzounopoulos2004cell, shin2006spatiotemporal, pawlak2008dopamine, sgritta2017hebbian}, and experiments have revealed diverse forms
of STDP.  However, no general theoretical framework exists, to our knowledge, to formally characterize the computations implemented by the myriad forms of this foundational plasticity rule.

Theoretical studies of STDP have addressed computational implications in specific learning scenarios. One line of research considers particular families of STDP rules and circuit architectures, and investigates the computations that they can implement, such as sequence prediction \cite{abbott1996functional}, Bayesian inference \cite{nessler2009stdp, nessler2013bayesian}, and input selectivity \cite{gutig2003learning, burkitt2004spike, meffin2006learning, gilson2009emergence, kerr2013delay, luz2016oscillations}. Another begins with a prescribed normative objective and derives STDP-like learning rules capable of optimizing it, such as for maximizing mutual information \cite{chechik2003spike, toyoizumi2007optimality}, timing sensitivity \cite{bell2004maximising}, cumulative reward \cite{rao2001spike} or minimizing firing entropy \cite{bohte2004reducing}. 
Yet, a means to infer the computational role of an arbitrary 
STDP rule remains lacking. 

Here, we develop a general theoretical framework for STDP by drawing upon the structures of game theory. 
We do so motivated by experimental studies of synaptic plasticity, which associated STDP with both cooperative and competitive interactions. Cooperation can arise when the coordinated activity of multiple inputs jointly shapes plasticity to achieve a collective function \cite{sjostrom2001rate, tazerart2020spike, bi1999distributed, cassenaer2007hebbian, mishra2016symmetric}. Conversely, the strengthening of some inputs can competitively suppress or weaken others \cite{lo1991activity, zhang1998critical, sajikumar2014competition, el2018locally}. Together, these observations suggest that cooperation and competition provide a fundamental organizing perspective on collective synaptic learning, making game theory a natural source of inspiration. 

Specifically, here, we study learning under general STDP rules in a canonical $n$-to-1 neural circuit. We show that STDP induces an explicit game among synaptic connections. Notably, the interaction between the STDP kernel and the postsynaptic potential (PSP) obeys a parity-selection rule: parity-matched STDP–PSP components generate cooperative interactions described by a potential game, whereas parity-mismatched components generate competitive interactions described by a zero-sum game. These two game components are independent, and play distinct computational roles. By establishing an exact mathematical equivalence between synaptic learning and a game, our framework explains how local STDP rules can collectively generate global computation.




\textit{\textbf{Game dynamics}}---In game theory, a game is specified by three basic elements: a set of players, indexed by subscripts $i \in \{1,2, ...n \}$; their strategies, represented here by the vector $s := (s_1, s_2, \ldots, s_n)^T$; and their payoff functions $u := \left(u_1(s), u_2(s), \ldots, u_n(s)\right)^T$. In playing the game, each player $i$ selects a strategy $s_i$ to maximize their own payoff $u_i$; for example, a bidder may choose a bid that maximizes their expected profit.

Despite the diversity of strategic scenarios, different games often share common underlying structures. In particular, all finite games admit a canonical decomposition into three components that capture aligned incentives, conflicting incentives, and incentives that are unaffected by unilateral strategic changes \cite{candogan2011flows}.

Motivated by this three-way interpretation, for natural systems that evolve continuously, we consider a differentiable game $G$ composed of a common-interest potential component $G^P$, a zero-sum interaction component $G^{ZS}$, and a nonstrategic component $G^N$. Specifically,
\begin{equation}
G := G^P + G^{ZS} + G^N,
\end{equation}
where the payoff vector $u$ is correspondingly decomposed into three payoff components,
\begin{align}
u &= u^P + u^{ZS} + u^N .
\end{align}
Each component is characterized by a distinct strategic property.

In the potential game, there exists a global game potential over all players’ strategies, $V(s)$, such that the change in any player’s payoff induced by a unilateral strategy change matches the corresponding change in the potential, i.e. $\partial u_i^P(s) / \partial s_i = \partial V(s) / \partial s_i, \forall i$. Thus, players in a potential game can be viewed as effectively cooperating to increase a common game potential, even though each acts in their own self-interest. Intuitively, individually self-interested behavior can still contribute to the collective welfare. By contrast, zero-sum games describe competitive interactions in which gains to some players necessarily come at the expense of others, with $\sum_i u^{ZS}_i (s) = 0, \forall s$. Such games generally do not admit a common potential that aligns the players’ incentives. Finally, in a nonstrategic game, the payoff of each player depends only on the strategies of the other players rather than on their own strategy, with $\partial u^N_i (s) / \partial s_i = 0, \forall i$, and therefore acts as a background component of the game.

Game dynamics describes how the players’ strategies evolve over time. When players adopt gradient-play dynamics, updating their strategies according to the gradients of their own payoffs, the resulting game dynamics becomes
\begin{align}
\dot s &\propto \nabla_s \odot u^P + \nabla_s \odot u^{ZS} + \nabla_s \odot u^N .
\label{eq: game w evolve}
\end{align}
where $\odot$ denotes element-wise multiplication. Combining this expression with the defining properties of the three components gives
\begin{align}
\nabla_s \odot u^P &= \nabla_s V(s), \\
\nabla_s \odot u^{ZS} &= (\partial_{s_1} u_1^{ZS}, \partial_{s_2} u_2^{ZS}, \ldots, \partial_{s_n} u_n^{ZS})^T, \\
\nabla_s \odot u^N &= 0. 
\end{align}

\textit{\textbf{Neural learning dynamics}}---At the neuroscientific level, we consider a minimal convergent circuit comprised of generalized linear spiking neurons, with a single postsynaptic neuron receiving input from $n$ presynaptic neurons. The $i^{\text{th}}$ input spike train is written as $x_i(t) = \sum_k \delta(t - t_i^k)$, and the output spike train of the postsynaptic neuron is $y(t) = \sum_k \delta(t - t_y^k)$, with $\delta$ denoting the Dirac delta function. As is common in the field, we assume that the input spike trains are jointly wide-sense stationary \cite{morrison2008phenomenological}.

The postsynaptic potential elicited by the $i^{\text{th}}$ input is the convolved signal
$z_i(t) = (h * x_i)(t) = \sum_k h(t - t_i^k),$
where $h$ is the PSP kernel. The total effect on postsynaptic neuron's potential is modeled as a weighted sum of these convolved inputs,
$q(t) = \sum_i w_i z_i(t)$, where $w_i$ is the signed synaptic weight of the $i^{\text{th}}$ input. The instantaneous firing rate is then given by a nonlinear function $\lambda(t) = \phi(q(t))$, from which the output spike train $y$ is generated.

STDP specifies the change in synaptic connectivity associated with a pair of presynaptic and postsynaptic spikes through a kernel $K(\tau)$, where $\tau=t_{\mathrm{post}}-t_{\mathrm{pre}}$ is the postsynaptic spike time relative to the presynaptic spike time, and $\int K(\tau)d\tau=0$. Accounting for all relevant spike pairs gives  
\begin{align}
    \dot w_i(t) = \int_{-\infty}^{\infty} K(\tau) y(t) x_i(t - \tau) d \tau
\end{align}

The resulting synaptic dynamics reflects both the fixed properties of the underlying system and fluctuations arising from individual realizations of the spike trains $x_i$ and $y$. Here, we focus on regimes in which the fixed component dominates the evolution and characterize it through the ensemble-averaged dynamics. Specifically, we consider an ensemble of systems with identical setup and initialization that differ only in their realizations of the spike trains.

Let $\langle \cdot \rangle$ denote the ensemble average. Applying the approximation $\langle x_i(t) y(t+\tau) \rangle \approx \langle x_i(t)\lambda(t+\tau) \rangle$ and retaining only the first-order term in the Taylor expansion of $\lambda$ around the reference potential $\langle q(t) \rangle$, we obtain the dynamics governing the evolution of the weight vector $w = (w_1, w_2,....w_n)^T$:
\begin{align}
    \dot w =  \ \phi'(\langle q \rangle) A  w ,
\label{eq: SNN w evolve}
\end{align}
with
\begin{align}
A_{ij} := & \int_{-\infty}^{\infty} \kappa(\tau)\, C_{x_i,x_j}(\tau) d\tau, 
\label{eq: Aij} \\
\kappa(\tau) := & \int_{-\infty}^{\infty} K(\rho) h(\rho-\tau) d\rho = (K * h^{\vee})(\tau) .
\end{align}

The dynamical matrix $A$ is determined jointly by the STDP-PSP interaction kernel $\kappa$ and the input-spike cross-correlation $C_{x_i,x_j}(\tau) := \left\langle  x_i(t) x_j(t+\tau)\right\rangle$. Since both are fixed, $A$ remains constant during throughout the evolution. The superscript $ ^{\vee}$ denotes time reversal, such that $h^{\vee}(t)=h(-t)$.

Decomposing the STDP and PSP kernels into even and odd parts yields a unique decomposition of their interaction kernel $\kappa$. The parity-matched terms contribute to $\kappa_{\text{even}}$, whereas the parity-mismatched terms contribute to $\kappa_{\text{odd}}$:
\begin{align}
\kappa(\tau) =& \ \kappa_{\text{even}}(\tau) + \kappa_{\text{odd}}(\tau),\\
\kappa_{\text{even}} :=& \ K_{\text{even}} * h_{\text{even}}^{\vee} + K_{\text{odd}} * h_{\text{odd}}^{\vee}, \label{eq: kappa_even}\\
\kappa_{\text{odd}} :=& \  K_{\text{even}} * h_{\text{odd}}^{\vee} + K_{\text{odd}} * h_{\text{even}}^{\vee}. \label{eq: kappa_odd}
\end{align}
This, in turn, yields a parity-based decomposition of the dynamical matrix:
\begin{align}
A =  A_{\text{even}} + A_{\text{odd}},
\end{align}

with 
\begin{align}
A_{\text{even}, ij} := & \int_{-\infty}^{\infty} \kappa_{\text{even}}(\tau)\, C_{x_i,x_j}(\tau) d\tau, 
\label{eq: Aeven} \\
A_{\text{odd}, ij} := & \int_{-\infty}^{\infty} \kappa_{\text{odd}}(\tau)\, C_{x_i,x_j}(\tau) d\tau, 
\label{eq: Aodd}
\end{align}
from eqs.(\ref{eq: Aij}, \ref{eq: kappa_even}, and \ref{eq: kappa_odd}).

 By construction, $A_{\text{even}}$ is symmetric, whereas $A_{\text{odd}}$ is antisymmetric. Consequently, they correspond to the gradient and solenoidal components, respectively, of a Helmholtz decomposition of $Aw$. (See Supplementary Material for the full derivation).



\textit{\textbf{Equivalence between game dynamics and neural learning}}---The above allows for the establishment of a direct correspondence between neural learning and the differentiable game constructed above. Specifically, learning in the canonical convergent circuit can be reinterpreted as a game (Fig. 1) in which each player $i$, corresponding to a presynaptic neuron, adjusts their strategy $w_i$, the synaptic strength, to maximize their payoff $u_i$, which encodes the computational role of that synapse. The payoff structure of this game incorporates both cooperation and competition, corresponding to different components of the STDP–PSP interaction.

Under this interpretation, the neural learning dynamics in Eq.~\eqref{eq: SNN w evolve} is equivalent to the strategy evolution of the differentiable game in Eq.~\eqref{eq: game w evolve}. Specifically, we obtain the following term-by-term correspondence:
\begin{equation}
\dot{w}
    \propto A_{\text{even}} w + A_{\text{odd}} w
     \longleftrightarrow 
    \dot{s}
    \propto \nabla_s V + \nabla_s \odot u^{ZS}.
\end{equation}
under the mapping
\begin{align}
s_i &= w_i \\
u_i^P &= V = \tfrac{1}{2} w^T A_{\text{even}} w,\ 
\label{eq: uP, V definition}
\\
u_i^{ZS} &= w_i \sum_j A_{\text{odd},ij} w_j,
\quad \sum_i u_i^{ZS} = 0, 
\label{eq: uH}
\\
u_i^N &= c_i. \nonumber
\end{align}
where $c_i$ is a constant independent of $w$. 

This correspondence yields that the parity-matched part of the STDP–PSP interaction defines the potential-game component and thereby, the cooperation among presynaptic players. Similarly, the parity-mismatched part defines the zero-sum component and thereby, the competition among players. The nonstrategic component captures weight-independent payoff offsets and acts as a constant background term in neural learning.

With this correspondence in place, we next investigate the computational roles encoded by the game payoff functions. Specifically, we ask: what do the neurons cooperate to achieve, and what do they compete for? We answer these questions by examining the two game components separately, identifying the distinct computations they implement, and then combining them to develop a unified interpretation of STDP learning.

\begin{figure}
    \centering
    \includegraphics[width=0.8\linewidth]{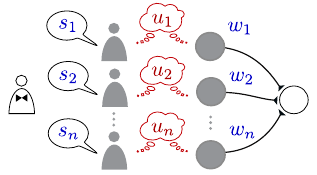}
    \caption{Illustration showing the proposed equivalence between game and synaptic learning in neural circuit.}
\end{figure}





\textit{\textbf{Computation by the potential-game component}}---To reveal its computational objective, we rewrite $V$ defined in Eq. \eqref{eq: uP, V definition} in a more interpretable form,  by exploiting the spectral properties of $\kappa_{\text{even}}(t)$. 
We show that $V$ 
is the difference between the  variances of two induced distributions 
obtained by applying specific synaptic filters to the input spike train: namely, filters derived from the parity matched STDP-PSP interaction.

The parity-matched STDP-PSP interaction kernel $\kappa_{\text{even}}(\tau)$ is even, and consequently, its Fourier transform, $\widehat{\kappa}_{\text{even}}(\xi)$, is real. Its positive and negative spectral parts, in turn, uniquely define two real filters, $g^{+}(\tau)$ and $g^{-}(\tau)$, respectively: 
\begin{align}
    \widehat{g^{+}}(\xi) &= \sqrt{\max (0, \,\widehat{\kappa}_{\text{even}}(\xi))} , \\  \widehat{g^{-}}(\xi) &= \sqrt{\max (0, -\widehat{\kappa}_{\text{even}}(\xi))}.
\end{align}

These filters can be used to express the potential $V = \frac{1}{2} w^TA_{\text{even}}w$ as the difference between two variances:
\begin{align}
V(w) = \frac{1}{2}\left(\text{Var}[w^T(g^{+}*x)] - \text{Var}[w^T(g^{-}*x)]\right), \label{eq: V-var}
\end{align}
where $x(t) = (x_1(t),\ldots,x_n(t))^T$ is the vector of inputs and
$g * x = (g*x_1,\ldots,g*x_n)$ denotes the vector of convolved inputs. Here
$\text{Var}[\cdot] := \left\langle \left(\cdot - \langle \cdot \rangle \right)^2 \right\rangle$ denotes signal variance (See Supplementary Material for the full derivation). 

From 
eq.(\ref{eq: V-var}), it becomes clear that, in general, potential-maximizing game dynamics are inherently unstable: if there exists a direction along which the variance of $g^{+}*x$ exceeds that of $g^{-}*x$, the potential is positive and drives $w$ to diverge. Otherwise, if no such direction exists, the variance difference is nonpositive and $w$ shrinks toward the zero vector, apart from possible neutral directions for which the difference vanishes. This instability accords with previous observations that STDP can drive synaptic weights collectively toward saturation or collapse \cite{chen2013heterosynaptic, volgushev2016partial}, rather than segregating them into a bimodal distribution of competing strong and weak subsets.

However, when $w$ is constrained to have a constant norm, 
as happens in biological neural networks through mechanisms such as homeostatic plasticity \cite{turrigiano1998activity, royer2003conservation}, the potential maximization  becomes equivalent to finding the direction along which the projected variances of the two induced distributions contrast most strongly. This corresponds precisely to the established algorithm of \textit{contrastive PCA}, whose goal is to extract informative signal from a target distribution that differs from a background distribution, for example, disease-related protein-expression variation against variation in healthy controls \cite{abid2018exploring}. Here, the neural circuit defines the target and background distributions as two distinct $n$-dimensional distributions obtained by applying the filters $g^{+}$ and $g^{-}$ to the input spike data.

A special case occurs when $\widehat{\kappa}_{\text{even}}(\xi)=1$ almost everywhere. Then $g^{+}$ reduces to the Dirac delta function and $g^{-}$ becomes zero, yielding ordinary PCA:
\begin{align}
V(w) = \frac{1}{2} \text{Var}[w^T x].
\end{align}




\textit{\textbf{Computation by the zero-sum component}}---This component is characterized by the payoff vector $u^{ZS}$, whose $i$-th entry, $u_i^{ZS}$, represents the individual payoff of player $i$ in competition. As shown in Eq.~\eqref{eq: uH}, this payoff function is governed by the parity-mismatched STDP-PSP interaction. 

After transforming to the frequency domain, the payoff can be re-expressed as a weighted integral of $\operatorname{Im}[\widehat{C}_{w_i x_i,v}(\xi)]$, the imaginary part of the cross-spectrum between the synaptic contribution $w_i x_i$ and the postsynaptic drive $v:=\sum_j w_j x_j$:
\begin{align}
u_i^{ZS} = 
 \frac{1}{\pi} \int_{0}^{\infty}  \ \text{Im}[\widehat{\kappa}_{\text{odd}}(\xi)] \  \text{Im} [\widehat{C}_{w_i x_i,v}(\xi)]d\xi. 
\end{align}
(See Supplemental Material for the derivation).

The cross-spectrum $\widehat{C}_{w_i x_i,v}(\xi)$ is complex-valued and encodes both the strength and phase lag of co-fluctuations between the contribution from one presynaptic neuron and the summed contribution of all neurons at frequency $\xi$. Its imaginary part, $\operatorname{Im}[\widehat{C}_{w_i x_i,v}(\xi)]$, isolates lagged synchronization from the in-phase component captured by the real part, a property commonly exploited by measures such as imaginary coherence (ImCoh) to characterize signal flow \cite{nolte2004identifying}. The magnitude of $\operatorname{Im}[\widehat{C}_{w_i x_i,v}(\xi)]$, therefore, reflects the strength of lagged coactivity between the presynaptic contribution and postsynaptic drive, while its sign indicates which signal leads in time.

The frequency domain kernel $\widehat{\kappa}_{\rm odd}(\xi)$ therefore imposes a preference criterion induced by the STDP–PSP interaction.
Its sign across frequencies determines the preferred direction of signal flow, whereas its magnitude determines the frequency bands that matter. Maximizing the payoff thus corresponds to strengthening synapses that promote directed signal flow within specified frequency ranges.

Furthermore, the antagonistic nature of the zero-sum game implies that, as one presynaptic player improves its alignment with the favored signal flow, this gain comes at the expense of the other players. This competition arises because the postsynaptic drive $v$ depends on all inputs: an adjustment to a particular weight $w_i$ can shift the phase of $v$, thereby altering its lagged synchronization with the other inputs. Our framework captures this competitive redistribution of connectivity through the zero-sum game structure.

\begin{figure}
    \centering
    \includegraphics[width=1.0\linewidth]{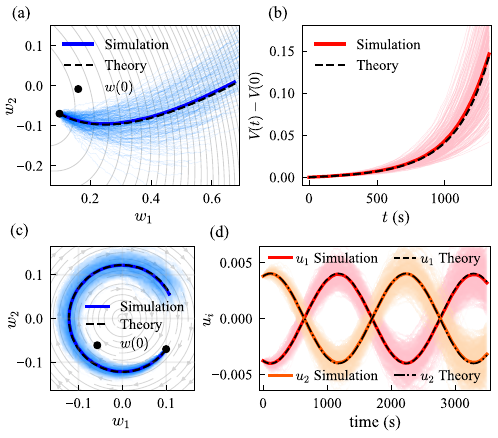}
    \caption{(a)–(b) Parity-matched simulations: (a) synaptic-weight dynamics (blue) and (b) corresponding potential dynamics (red). (c)–(d) Parity-mismatched simulations: (c) synaptic-weight dynamics (blue) and (d) payoff dynamics of the two players (red and orange).  In all panels, individual trajectories are shown in light colors, their ensemble mean in darker colors, and theoretical predictions in black.  }
\end{figure}

\textit{\textbf{Parity-matched simulations}}---To validate our theoretical prediction that parity-matched and parity-mismatched STDP–PSP interactions give rise to potential-game and zero-sum dynamics, respectively, we simulate an ensemble of 2-to-1 SNNs, for which the two-dimensional weight vector $(w_1, w_2)$ can be directly visualized. We first consider the parity-matched case, where both the STDP and PSP kernels are chosen to be even functions, with
\begin{align}
K_e^{\text{sim}}(\tau)
&=
\eta \left(1-\frac{\tau^2}{\sigma^2}\right)
e^{-\frac{\tau^2}{2\sigma^2}}, \\
h_e^{\text{sim}}(\tau)
&=
e^{-\frac{\tau^2}{2\sigma^2}}.
\end{align}
The parameter $\eta$ is a plasticity scaling factor that controls the learning rate, while $\sigma$ sets the temporal width of the kernels. The input spike trains are chosen as 
two time-shifted copies of the same Poisson spike train with firing rate 
$r$, separately by a delay of $\sigma$. Their correlation functions are:
\begin{align}
C_{x_1,x_1}^{\text{ sim}}(\tau)
&=
C_{x_2,x_2}^{\text{ sim}}(\tau)
=
r^2+r\delta(\tau), \\
C_{x_1,x_2}^{\text{ sim}}(\tau)
&=
r^2+r\delta(\tau-\sigma), \\
C_{x_2,x_1}^{\text{ sim}}(\tau)
&=
r^2+r\delta(\tau+\sigma).
\end{align}
(See Supplemental Material for details of the simulation setup.)
Both the ensemble-averaged weight trajectories (Fig. 2(a)) and the increase in game potential (Fig. 2(b)) agree with the theoretical predictions.







\textit{\textbf{Parity-mismatched simulations}}---To validate the prediction that this component 
generates zero-sum game dynamics, we simulated the same circuit structure with the same input statistics as in the potential-game case. The only change was to replace the STDP kernel with an odd function, leaving only the parity-mismatched component:
\begin{align}
K_o^{\text{sim}}(\tau)
&=
\eta \frac{\sqrt{5}}{2}\frac{\tau}{\sigma}
e^{-\frac{\tau^2}{2\sigma^2}}.
\end{align}

The ensemble-averaged weight dynamics exhibit cyclic competitive behavior (Fig. 2(c)), while the payoffs of the two presynaptic players are anticorrelated over time (Fig.2(d)). Both observations agree with the theoretical predictions (See Supplemental Material for details of the simulation setup).

We note that, although the simulation displays a closed cycle, such cyclic competition would be interrupted in a biological circuit by sign constraints arising from Dale’s principle \cite{eccles1954cholinergic}. An excitatory output cannot become inhibitory, so the trajectory eventually encounters a hard boundary rather than completing the cycle. In game-theoretic terms, this boundary corresponds to a decisive victory for one player and the elimination of a bankrupt opponent.



\textit{\textbf{Mixed-parity simulations}}---Finally, we simulate the general case in which both the STDP and PSP kernels contain even and odd components. The resulting neural learning dynamics combines a cooperative potential-game component with a competitive zero-sum component. We use the same circuit setup and input statistics as before, but parameterize the STDP and PSP  kernels by $(\alpha, \beta)$ (Fig. 3(a)), which control the relative contributions of their even and odd components:
\begin{align}
K_\alpha^{\text{sim}}(\tau)
&=
\cos\alpha\,K_e^{\text{sim}}(\tau)+\sin\alpha\,K_o^{\text{sim}}(\tau), \\
h_\beta^{\text{sim}}(\tau)
&=
\cos\beta\,h_e^{\text{sim}}(\tau)+\sin\beta\,h_o^{\text{sim}}(\tau),
\end{align}
where $K_e^{\text{sim}}, K_o^{\text{sim}}, h_e^{\text{sim}}$ are defined in the previous simulations, and 
\begin{align}
h_o^{\text{sim}}(\tau)
&=
\frac{2}{\sqrt{5}}\frac{\tau}{\sigma}
e^{-\frac{\tau^2}{2\sigma^2}} .
\end{align}

Under this setup, the parity-matched and parity-mismatched components of the dynamical matrix are (See Supplemental Material for the derivation)
\begin{equation}
A_{\rm even}^{\text{pred}}
=
\eta \frac{1}{4} r\sqrt{\pi}\sigma \cos (\alpha - \beta)
\begin{pmatrix}
2
&
e^{-1/4}
\\
e^{-1/4}
&
2
\end{pmatrix},
\end{equation}
and
\begin{equation}
A_{\rm odd}^{\text{pred}}
=
\eta \frac{\sqrt{5}}{4}r\sqrt{\pi}\sigma e^{-1/4}
\sin(\alpha-\beta)
\begin{pmatrix}
0 & 1\\
-1 & 0
\end{pmatrix}.
\end{equation}

These two components represent the cooperative and competitive contributions to the learning dynamics. Their relative magnitudes determine whether learning is cooperation-dominated or competition-dominated. We therefore define the \textit{ratio of competition} from their Frobenius norms:
\begin{equation}
R(\alpha,\beta)
:=
\frac{\|A_{\rm odd}(\alpha,\beta)\|_F^2}
{\|A_{\rm even}(\alpha,\beta)\|_F^2
+
\|A_{\rm odd}(\alpha,\beta)\|_F^2}.
\end{equation}

This yields the following analytical prediction as a function of $(\alpha, \beta)$:
\begin{equation}
R^{\text{pred}}(\alpha,\beta)
=
\frac{
5\sin^2(\alpha-\beta)
}{
(1+4e^{1/2})\cos^2(\alpha-\beta)
+
5\sin^2(\alpha-\beta)
}.
\end{equation}

\begin{figure}
    \centering
    \includegraphics[width=1.0\linewidth]{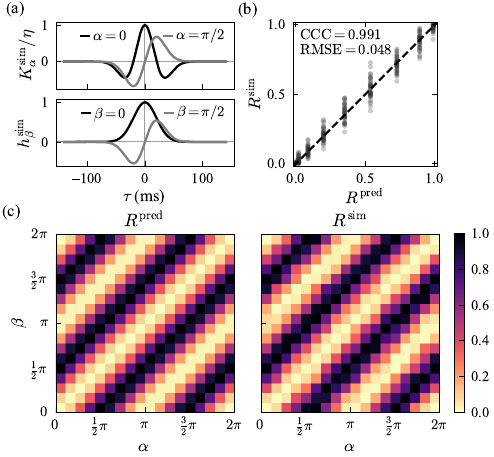}
    \caption{ Simulation results for the mixed-parity scenario. (a) Even and odd components of the normalized STDP and PSP kernels. (b) Simulated versus theoretical competition ratios across the $(\alpha,\beta)$ parameter space. Agreement is quantified by Lin's concordance correlation coefficient (CCC) and root-mean-square error (RMSE). (c) Corresponding heatmaps of the theoretical and simulated competition ratios.}
\end{figure}

We simulate ensembles of SNNs across a range of parameter pairs $(\alpha, \beta)$. The measured \textit{ratio of competition} closely matches the theoretical prediction (Figs.~3(b) and 3(c)), with Lin’s concordance correlation coefficient of 0.991 (See Supplemental Material for details of the simulation setup).




\begin{figure}
    \centering
    \includegraphics[width=1.0\linewidth]{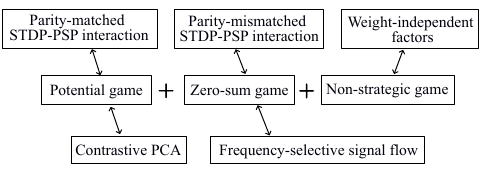}
    \caption{Summary of the theoretical framework.}
\end{figure}

\textit{\textbf{Discussion}}---In this work, we introduced a framework that establishes a mathematical equivalence between learning dynamics in a spiking neural network and strategy evolution in a game (Fig. 4). In a manner reminiscent of Minsky’s “Society of Mind” \cite{minsky1986society}, learning emerges from interactions among presynaptic synapses viewed as individual players. The parity decomposition reveals that these interactions contain both potential-game and zero-sum components (Fig. 4). The potential-game component supports objective-driven computation through cooperative synaptic interactions, giving rise to contrastive principal component analysis. The zero-sum component instead produces competition over the strength of directed signal transmission within specific frequency bands. These two components coexist, allowing the same learning system to perform cooperative and competitive computations simultaneously.


Whereas previous works have utilized some elements similar to those in our approach, such as STDP–PSP convolution or symmetric–antisymmetric decomposition \cite{gilson2012spectral, sprekeler2007slowness}, they have provided only partial explanations of the underlying computations by showing that neural learning resembles slow-feature analysis (SFA) or PCA in more restrictive cases.
Our work extends these past lines of research by providing a game-theoretic mathematical formalization that applies to a much broader range of STDP and PSP kernels and unravels the precise algorithmic computation implemented by neural learning.
Moreover, by viewing the mixture of different components as an integrated system, our framework reveals how both the computational role and the learning dynamics are systematically shaped by the parity structure of the STDP–PSP interaction.

From the perspective of game theory, the general evolution of a game need not maximize the collective welfare of its participants \cite{roughgarden2002bad, stewart2014collapse, mazumdar2020gradient}, 
motivating principles such as incentive alignment \cite{maskin2008mechanism, balduzzi2018mechanics} in the theory of mechanism design for economic and social systems. From this viewpoint, biological processes including homeostatic plasticity \cite{turrigiano2004homeostatic}, neuromodulation \cite{nadim2014neuromodulation}, and metaplasticity \cite{abraham2008metaplasticity} may be interpreted as mechanisms that align local learning incentives in ways that support global computation. This interpretation suggests that tools from game theory may be useful for analyzing how biological systems coordinate local plasticity rules to achieve system-level computational functions.

Our framework also suggests a possible mechanism for representational drift \cite{rule2019causes, driscoll2022representational}, whereby the tuning of individual neurons changes over time while task-relevant information remains stably encoded at the population level, supporting consistent behavior. The cooperative game component drives the system toward stable solutions in the potential landscape, allowing the population-level computation to converge. In contrast, the competitive zero-sum component can sustain non-convergent dynamics, as individual players continually adapt in response to one another. The coexistence of these two components therefore provides a natural separation between computational stability and representational stability: the population-level computation may remain stable while individual synaptic weights, and consequently neuronal tuning, continue to evolve. This predicts that representational drift may arise intrinsically from parity-mismatched STDP–PSP interactions, rather than solely from noise or learning from new experiences.

Finally, our results demonstrate that both the computational regime and the qualitative learning dynamics can be controlled simply by altering the temporal structure of the STDP and PSP kernels, as established analytically and further illustrated by progressively varying these kernels in simulations. This direct link between temporal kernel structure and emergent computation may help explain how diverse plasticity rules across cortical and subcortical structures support distinct computational functions \cite{bi1998synaptic, lu2007spike, fino2010distinct}, while also providing a principled basis for designing STDP rules that implement desired computations and dynamical regimes. Beyond individual plasticity rules, our results point to contrastive PCA as a promising computational primitive for building larger artificial neural systems with stable population-level computation. Architectures composed of such modules, rather than conventional perceptron-like units, may offer a biologically grounded, unsupervised alternative for constructing multilayer artificial neural systems.



\bibliography{refs_neuralgame}

\end{document}